\documentclass[aps,twocolumn,pre,showpacs,eqsecnum]{revtex4}

\usepackage{graphpap}
\usepackage[dvips]{graphicx}
\usepackage[dvips]{graphics}
\usepackage{color}

\begin{document}

%--------------------------------------------------
\title{Tunable Graphene Single Electron Transistor}
%--------------------------------------------------

 \author{C. Stampfer\footnote{Corresponding author, e-mail: stampfer@phys.ethz.ch}, 
 E. Schurtenberger, F. Molitor, J. G\"uttinger, T. Ihn, and K. Ensslin}

\affiliation{Solid State Physics Laboratory, ETH Zurich, 8093 Zurich, Switzerland}

\date{ \today}
 
\begin{abstract}

We report electronic transport experiments on a graphene single electron transistor.
The device consists of a graphene island connected to source and drain electrodes via two narrow graphene constrictions. It is electrostatically tunable by three lateral graphene gates and an additional back gate. The tunneling coupling is a strongly nonmonotonic function of gate voltage indicating the presence
of localized states in the barriers.
We investigate energy scales for the tunneling gap, the resonances in the constrictions and for the Coulomb blockade resonances. 
From Coulomb diamond measurements in different device configurations (i.e. barrier configurations)  we extract a charging energy of~$\approx$~3.4~meV and estimate a characteristic energy scale for the
constriction resonances of~$\approx$~10~meV.
% and peak spacing fluctuations of approximately 20\%. %For increasing temperatures we detect a peak broadening and a transmission increase of the nanostructured graphene barriers. 

\end{abstract}

\maketitle

\newpage

The recent discovery of graphene~\cite{nov04,gei07}, filling the gap between quasi 1-dimensional (1-D) 
nanotubes
and 3-D graphite 
makes truly 2-D crystals accessible and links solid state devices to molecular electronics~\cite{joa00}. 
Graphene, which exhibits unique electronic properties including massless carriers near the Fermi level and potentially weak spin orbit and hyperfine couplings~\cite{min06,tom07} has been
proposed to be a promising material for spin qubits~\cite{tra07}, high mobility electronics~\cite{che07,han07} and it may have the potential to contribute to the downscaling of state-of-the-art silicon technology~\cite{ieo04}. 
The absence of an energy gap in 2-D graphene and phenomena related to Klein tunneling~\cite{dom99,kat06} make it hard to confine carriers electrostatically and to 
control transport on the level of single particles. 
However, by
focusing on graphene nanoribbons, which are known to exhibit an effective transport gap~\cite{che07,han07,sol07,dai08} this limitation can be overcome. It has been shown recently that such a transport gap allows
to fabricate well tunable graphene nanodevices~\cite{sta08,pon08,mia07}. %\cite{no}. 
Here we investigate a fully tunable single electron transistor (SET) that consists of a width modulated graphene structure exhibiting spatially separated transport gaps. SETs consist of a conducting island connected by tunneling barriers to two conducting leads. Electronic transport through the device can be blocked by Coulomb interaction for temperatures and bias voltages lower than the characteristic energy required to add an electron to the island~\cite{kou97}.
\newline
The sample is fabricated based on single-layer graphene flakes obtained from mechanical exfoliation of bulk graphite. These flakes are deposited on a highly doped silicon substrate with a 295~nm silicon oxide layer~\cite{nov04}. 
Electron beam (e-beam) lithography is used for patterning the isolated graphene flake by subsequent Ar/O$_2$ reactive ion etching.
Finally, an additional e-beam  and lift-off step is performed to pattern Ti/Au (2~nm/50~nm) electrodes.
For the detailed fabrication process and the single-layer graphene verification we refer to Refs.~\cite{sta08,fer06,dav07a}. 
Fig.~1a shows a scanning force micrograph of the investigated device. Both the metal electrodes and the graphene structure are highlighted.  
In Fig. 1b, a schematic illustration of the fabricated graphene SET device is shown. Source (S) and drain (D) contacts connect via 50~nm wide constrictions to the graphene island. The two constrictions are separated 
 by $\approx$~750~nm and the island has an area $A\approx 0.06~\mu$m$^2$ (see Figs.~1a,b).
 \begin{figure}[h]\centering
\includegraphics[draft=false,keepaspectratio=true,clip,%
                   width=0.98\linewidth]%
                   {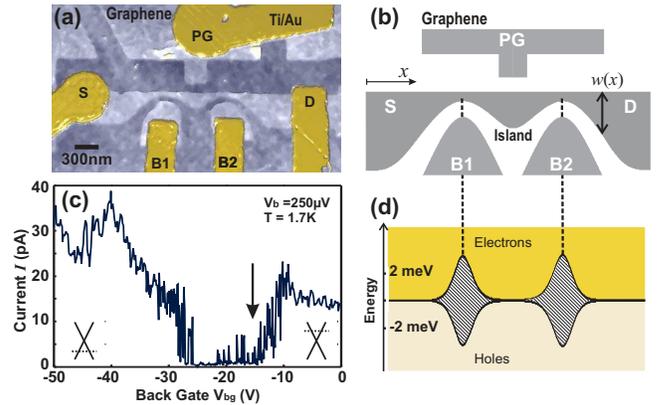}                   
\caption[FIG1]{(color online)
(a) Scanning force microscope image of the investigated graphene single electron transistor (SET) device, where the graphene structure and the metal electrodes are highlighted. The minimum feature size is approx.~50~nm.
(b) Schematic illustration of the tunable SET device with electrode assignment.
(c) Low bias back gate trace for $V_{b1}=V_{b2}=V_{pg}=0$~V. The resolved transport gap separates between hole and electron transport.
(d) Effective energy band structure of the device as depicted in Fig.~1b. The tunnel barriers exhibit an effective energy gap of approx. 6.5~meV. For more information of this model see text.} 
\label{trdansport}
\end{figure}
\begin{figure*}[t]\centering
\includegraphics[draft=false,keepaspectratio=true,clip,%
                   width=0.94\linewidth]%
                   {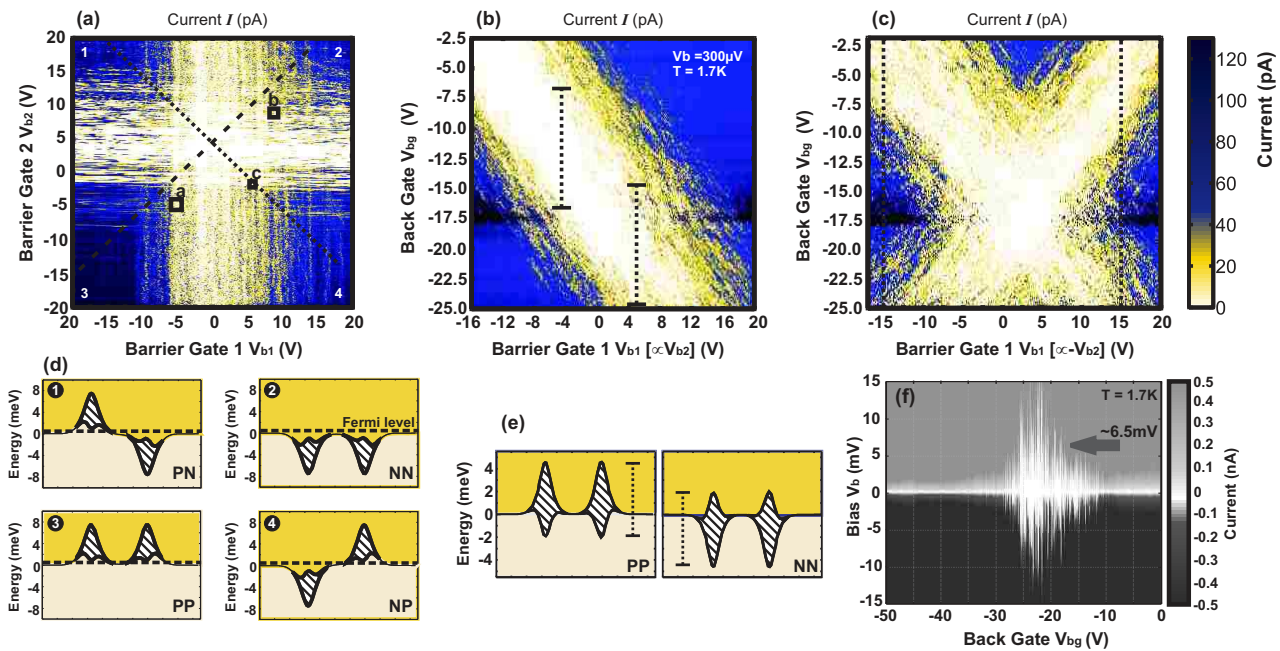}                   
\caption[FIG1]{
(color online)
Transport as function of the barrier gate potentials $V_{b1}$, $V_{b2}$ and the back gate at small bias voltages. 
(a)~Source-drain current plotted as function of $V_{b1}$ and $V_{b2}$ for constant back gate ($V_{bg}=-15$~V; see arrow in Fig.~1c). Here, both individual gaps can clearly be seen. The labels a-c are related to the corresponding close-ups shown in Fig.~3.
(b)~Symmetric barrier gate voltages $V_{b1}=V_{b2}-5$~V~\cite{com01} as function of a varying 
back gate voltage at $V_{b}=300~\mu$V. The white areas correspond to suppressed current. 
(c)~shows the same for antisymmetric barrier gate voltages $V_{b1}=-V_{b2}-5$~V, where both 
transport gaps are clearly visible. Please note also the gap homogeneity as function of the back gate. 
 (d)~Schematic illustration of the barrier configurations explaining the different transport regimes shown in Fig.~a. 
(e)~Schematic illustrations for symmetric tuning of the tunnel barriers corresponding to Fig.~b.
(f)~Source-Drain current as function of bias and back gate voltage (all other gates have been grounded). The measured effective energy gap agrees reasonable well with the model calculation (see arrow). For more details see text.
}
\end{figure*} 
In order to tune the two tunneling barriers and the island electrostatically and
independently, three lateral graphene gates~\cite{mol07} have been fabricated
closer than 100~nm to the active graphene
structure (see Fig.~1a). 
These are the two barrier gates B1 and B2,
and the plunger gate PG (Fig.~1b).
The additional highly doped silicon substrate is used as a back gate (BG) to adjust the overall Fermi level ($E_F$).

All measurements have been performed in a variable temperature $^4$He cryostat at a 
base temperature of T$\approx$~1.7~K and the sample was heated to 135$^{\circ}$C in vacuum for 12~h before cooling down.
We have measured the two-terminal conductance through the graphene SET device by applying a symmetric DC bias voltage $V_{b}$ while measuring the current through the SET device with a resolution better than 10~fA. For differential conductance measurements a small AC bias, $V_{b,ac}=50~\mu$V has been superimposed on $V_{b}$ and the differential conductance has been measured with lock-in techniques.

At small bias ($V_b = 250~\mu$V$~< 4 k_B T$) strong current suppression  
is observed at $-25$~V$~<~V_{bg}~<~-15$~V,~as shown in Fig.~1c.
This suppression is in agreement with earlier studies of graphene nanoconstrictions~\cite{che07,han07}.
It can be interpreted as a transport gap forming around the back gate voltage where the system
is charge neutral. Hole transport occurs at $V_{bg}<-25$~V, electron transport at $V_{bg}>-15$~V.

Measurements for varying back gate voltage (Fermi level) and bias voltage 
allow to estimate the size of the transport gap as shown in Fig.~2f.
A value of the order of $10$~meV is found. 
However, the strong modulation of the current shows, that localized states lead to 
strong transmission resonances. Therefore we refer in the following to an "effective" energy gap or a 
transport gap.  

The geometric design of our structure (see Fig. 1a) gives local electrostatic access to the constriction regions. Fig. 2a shows a measurement of the current where the voltages $V_{b1}$ and $V_{b2}$ on the two barrier gates B1 and B2 have been independently tuned while the back gate voltage was kept fixed at $V_{bg} = -15$~V. A vertical and a horizontal stripe of suppressed current is observed. This observation indicates that transport through each of the two constrictions is characterized by a transport gap which can be individually tuned with the respective barrier gate. For example, keeping $V_{b1} = -20$~V constant and sweeping $V_{b2}$ from -20~V to + 5~V keeps constriction~1 conducting well while constriction~2 is tuned from large conductance to very low conductance (into the transport gap). The capacitive cross talk from B1 to constriction~2, and from B2 to constriction~1 is found to be smaller than 2$\%$.

These measurements suggest that the energy diagram shown in Fig.~1d is a useful description of the data. In this figure, high (electron) and low (hole) energy states are separated by two solid lines. Outside the constriction regions these lines are degenerate and represent the energy of the charge neutrality point in graphene. In the constriction regions the two lines are energetically separated indicating the observed effective energy (transport) gap $E_g$ by hatched areas. As a result of the lack of an energy gap of the two-dimensional graphene material, the exact shape of the effective $E_g(x)$ ($x$ is the transport direction) is given only by lateral confinement, i.e., by the variation of the width $w(x)$ along the device. We assume that electron-hole symmetry holds in the confined geometry and therefore plot an effective conduction band edge at $+E_g(x)/2$, and an effective valence band edge at $-E_g(x)/2$.

%
%-----------------------------------------------------------
It is known from earlier experiments~\cite{che07,han07} that graphene nanoribbons
(or constrictions) exhibit an effective energy gap.
For ribbons of width $w < 20$~nm the size of this gap scales according to $E_g=\hbar v_F/w$, where $v_F=10^6$~m/s is the Fermi velocity. The energy gap for nanoribbons wider than 20~nm can be reasonably well described by $E_g(w)=a/w \exp(-bw)$~\cite{sol07}, where $a=1$~eV$\times$nm and $b=0.023~$nm$^{-1}$ are constants extracted from fits of the experimental data in Ref.~\cite{han07}. Within this model the width $w(x)$ of our graphene structure translates to an effective transport band structure exhibiting two tunnel junctions with barrier height $E_{g,b} = 6.5$~meV and an almost gap free island ($E_{g,i} = 85~\mu$eV) as shown in Fig. 1d. According to the model the SET is expected to be operational in the regime of $\left|E_F\right|<E_{g,b}/2$. The measured transport gap agrees reasonably well with the modeled barrier height, as indicated by the arrow in Fig.~2f.

The local electrostatic influence of the gate electrodes can be incorporated into this heuristic description as a local shift of the energy of the charge neutrality point described by smooth characteristic potentials $\phi_{i}(x)$ ($i=b1,b2,pg,bg$) which may be derived from purely electrostatic considerations. While $\phi_{bg}(x)$ is independent of $x$, $\phi_{b1}(x)$ and $\phi_{b2}(x)$ are peaked at the respective constrictions, and $\phi_{pg}(x)$ is peaked within the island. For creating the schematic figures in this paper [Figs. 1(d), 2(d), (e)] we have used a convenient peaked $\phi_i(x)$ function (the shape of which is irrelevant for this simple discussion) with peak heights compatible with lever arms extracted from the experiment (see below).

Having established a heuristic energy diagram describing our sample we now return to the discussion of the measurement in Fig. 2a which is facilitated by the diagrams in Fig.~2d. In this measurement 
$V_{bg} = -15$~V. From Fig. 1c we deduce that the Fermi energy in the contacts of the structure lies within the conduction band, as indicated by the horizontal dashed lines in the four drawings in 
Fig.~2d. The four drawings represent energy diagrams corresponding to the four corners of Fig.~2a as indicated by the white numbers. In corner 2 transport takes place in the conduction band throughout the whole structure. In corner 1 (4) transport occurs in the conduction band in the right (left) part of the structure. The left (right) constriction is traversed via states in the valence band. The situation is even more complex in corner 3, where the Fermi energy cuts both barrier regions in the valence band. Although these situations imply two or even four p-n-like transitions along the structure, no distinctive features are observed in our measurements. This may be a manifestation of the suppression of backscattering due to Klein tunneling.

Figs.~2b and 2c demonstrate the consistency of our heuristic model with the experimental observations. Fig.~2b shows the current measured as a function of $V_{bg}$ and $V_{b1}$, with $V_{b2}$ being simultaneously swept such that $V_{b2} = V_{b1} + 5$V (see dashed line in Fig.~2a). In this way the barrier regions are simultaneously shifted up or down (see Fig.~2e). Fig. 2b shows that the transport gap measured as a function of the back gate is shifted correspondingly, with $\Delta V_{bg}/\Delta V_{b1,2} \approx 0.9$. 

Fig. 2c shows the current measured as a function of $V_{bg}$ and $V_{b1}$, with $V_{b2}$ being simultaneously swept such that $V_{b1} + V_{b2} = 5$~V (see dotted line in Fig.~2a). For $V_{b1} = \pm15$~V (vertical dashed lines in Fig.~2c) the position of the gaps in energy correspond to diagrams 1 and 4 in Fig. 2d. In these two cases, sweeping the back gate allows to probe the two spatially separated transport gaps individually.

If we focus on a smaller voltage scale much more finestructure in the $V_{b1}-V_{b2}$ parameter plane appears, as shown in Fig.~3. Subplots 3a-c are different close ups of Fig.~2a (see 
black labeled boxes therein).
\begin{figure}[t]
\centering
\includegraphics[draft=false,keepaspectratio=true,clip,%
                   width=0.95\linewidth]%
                   {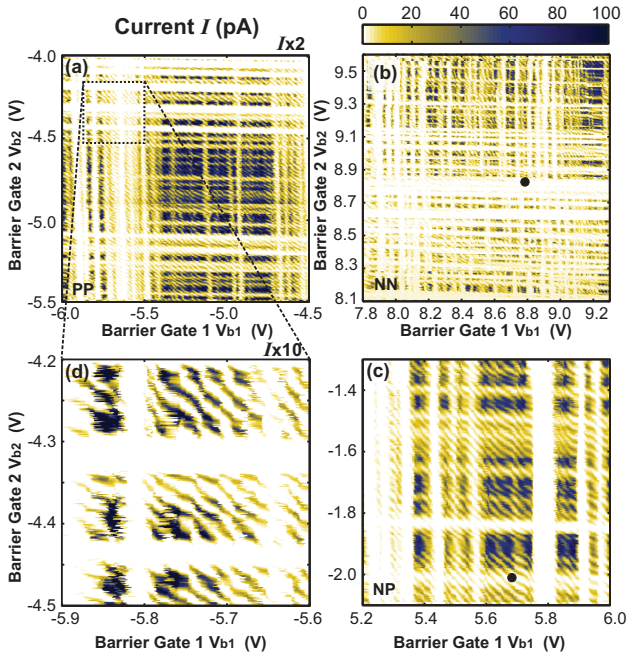}                   
\caption[FIG1]{
(color online) Source-drain current through the graphene SET as function of the barrier
gates $V_{b1}$ and $V_{b2}$ for constant bias $V_b=300~\mu$V and back gate $V_{bg}=-15$~V.
(a-c) are close-ups of Fig.~2a (as indicated therein by labeled boxes), showing transport in the PP (a), NN (b) and NP (c) regime. Ontop the horizontal and vertical transmission modulations we observe (diagonal) Coulomb blockade resonances. This is best seen in Fig.~d, which is a close-up of Fig.~a. 
In Figs.~a and d the current has been multiplied by a factor 2 and 10, respectively, to meet the color scale shown above b.
} 
\label{trdansport}
\end{figure}
\begin{figure}[t]
\centering
\includegraphics[draft=false,keepaspectratio=true,clip,%
                   width=0.85\linewidth]%
                   {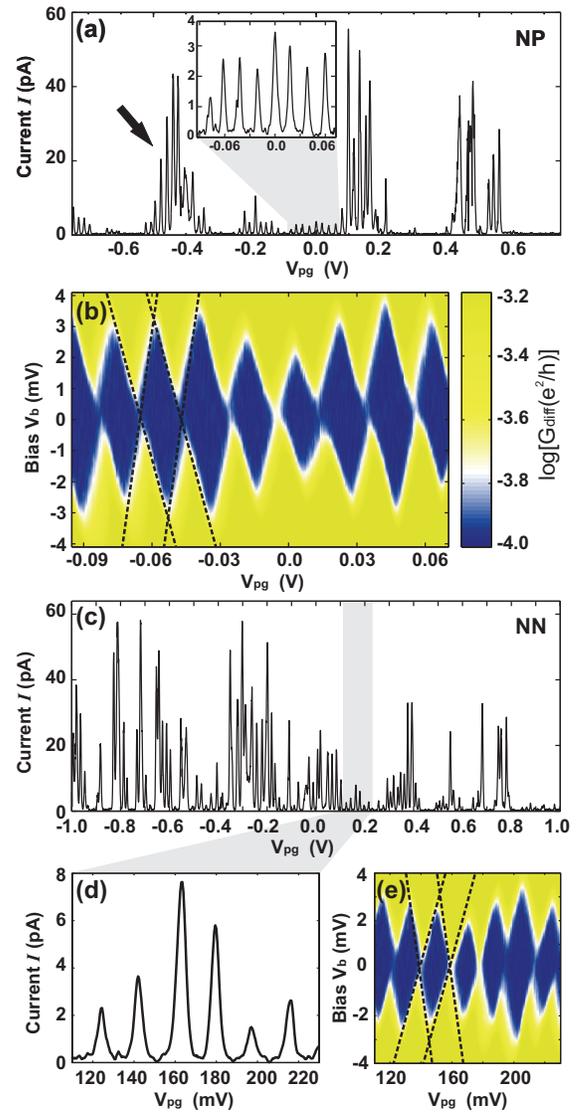}                   
\caption[FIG1]{
(a) Source drain current as function of the plunger gate voltage $V_{pg}$ 
at fixed back gate and barrier gates in the NP regime ($V_{bg}=-15$~V, $V_{bg1}=5.67$~V, and $V_{bg2}=-2.033$~V).
The inset (close up) clearly shows Coulomb peaks
(b) Corresponding Coulomb diamonds in differential conductance $G_{diff}$, represented in a logarithmic color scale plot (dark regions represent low conductance). A DC bias $V_{bias}$ with a small AC modulation (50~$\mu$V) is applied symmetrically across the dot and the current through the dot is measured. %Differential conductance has been directly measured by a Lock-in amplifier. 
(c) Coulomb resonances ontop and nearby strong transport modulations in the NN regime ($V_{bg}=-15$~V, $V_{bg1}=8.79$~V, and $V_{bg2}=8.85$~V). (d) shows a close up highlighting Coulomb peaks and (e) shows the corresponding Coulomb diamond measurements.
The color scale is adapted from Fig.~b.
 } 
\label{trdansport}
\end{figure}
Although Figs.~3a-c show the current in three different regimes the transport
characteristics do not differ significantly. Here, we distinguish
between the PP (Fig.~3a), NN (Fig.~3b) and the NP (Fig.~3c) regime, depending on either having the tunnel barriers (according to B1 and B2) shifted down (N) or up (P).
%As a consequence e.g. we do not expect any p-n-like junctions for the NN (or PP) regime (see e.g.
%right panel of Fig.~2e).
%
We observe in all regimes (Fig.~3) 
sequences of horizontal and vertical stripes of suppressed current and current resonances.
Their direction in the $V_{b1}$-$V_{b2}$ plane indicates that their physical origin has to
be found within constriction~1 (vertical stripes) or constriction~2 (horizontal stripes).
A blow-up of a small region in Fig.~3a is shown in Fig.~3d. The current exhibits even finer
resonances which are almost equally well tuned by both constriction gates. We therefore
attribute these resonances to states localized on the island between the barriers. It
will be shown below that these resonances occur in the Coulomb blockade regime of the island.
We attribute the deviations from perfectly straight diagonal lines to the presence of 
rough edges and inhomogeneities within the graphene island which has dimensions (slightly) larger than the elastic mean free path.

This characteristic pattern (Fig.~3d) can be found within a large $V_{b1}$-$V_{b2}$ parameter range within the regime where the two barrier gaps cross each other (i.e. the inner bright part of Fig.~2a).

So far we mainly focused on the barriers and in the following we concentrate on the charging of the island itself.
We fix the barrier gate potentials ($V_{b1}$ and $V_{b2}$) 
either in the NN regime or in the NP regime in  
order to study Coulomb blockade.
Fig.~4a shows sharp conductance resonances with a characteristic period of about 20~mV ($V_{b1}$=5.570~V, and $V_{b2}$=-2.033~V are fixed). Their amplitude is modulated on a much larger voltage scale of about 200~mV by the transparency modulations of the constrictions (cf. Fig.~3d). These resonances in the narrow graphene constrictions can significantly elevate the background of the Coulomb peaks (see e.g. black arrow). 
The inset of Fig.~4a
confirms that transport can also be completely pinched off between Coulomb blockade peaks. 
Corresponding 
Coulomb diamond measurements~\cite{kou97}, i.e., measurements of the differential conductance ($G_{diff}=dI/dV_b$) as function of bias voltage $V_{b}$ and plunger gate voltage $V_{pg}$ are shown in Fig.~4b. Within the swept plunger gate voltage range no charge rearrangements have been observed and the peak positions were stable over more than 10 consecutive plunger gate sweeps.

In Fig.~4c we show conductance resonances, which have been measured within the NN regime (for fixed $V_{b1}=8.79$~V and $V_{b2}=8.85$~V, see Fig.~3b).
The $V_{pg}$ range shown here is wider than in Fig.~4a. Again we observe (i) strong
transport modulations on a $V_{pg}$ scale of about 100~mV, which originate from resonances within the barriers and (ii) Coulomb peaks on a $V_{pg}$ scale of about 20~mV which are blown up
in Fig.~4d.
The corresponding Coulomb diamond measurements (Fig.~4e) 
are similar to those measured in the NP regime (Fig.~4a).
The Coulomb peaks (Fig.~4d and inset in Fig.~4a) and the Coulomb diamonds are not very sensitive to 
the tunnel barrier regime, although in one case a p-n-like junction should be present, whereas in the other case a more
uniform island is expected.

From the extent of all the diamonds in bias direction we estimate the average charging energy of the graphene single electron transistor operated in both regimes to be $E_C \approx 3.4$~meV.
This charging energy corresponds to a sum-capacitance of the graphene island $C_{\Sigma}=e^2/E_C~\approx~47.3$~aF, whereas the extracted back gate capacitance $C_{bg}~\approx~18$~aF is higher than the purely geometrical parallel plate 
capacitance of the graphene island $C=\epsilon_0 \epsilon A/d~\approx~7.4$~aF. This is related to the fact that 
the diameter of the graphene island ($\sqrt{A}$) is approximately the same as the gate oxide thickness $d$~\cite{sta08,ihn04}.

\begin{table}[t]
\begin{center} {\footnotesize
\begin{tabular}{|c||c|c|c|c|c|c|}
\hline
${ }$ & {BG} &
{PG} & {B1} & B2 & Source (S) & Drain (D) \\
\hline\hline
Capacitance (aF)  &    18.0  &  6.9  &   6.0 (5.5) & 5.0 & 1.8 (10.1) & 9.6 (1.8)\\\hline
Lever arm  &    0.38  & 0.15  &   0.13 (0.12)  &  0.1  & 0.04 (0.21) & 0.20 (0.04) \\
\hline
\end{tabular} }
\end{center}
\caption{Capacitances and lever arms of the different gate electrodes, including source and drain 
contacts, with respect to the graphene island. Most values are independent from the measurement regime, NN or NP. If there is a difference the NP value is given and the NN value is put in brackets.}
\label{turns}
\end{table}

The lever arms, and the electrostatic couplings of the electrodes to the graphene island
do not change significantly between the NN, PP (not shown) and the NP regime. Thus, the lever arm of the
plunger gate is $\alpha_{pg}~\approx~C_{pg}/C_{\Sigma} \approx 0.15$ ($C_{pg}~\approx~6.9$~aF), whereas the electrostatic
coupling to the other gates were determined to be $C_{b1}~\approx~5.5 - 6.0$~aF and $C_{b2}~\approx~5.0$~aF.
\begin{figure}[t]
\centering
\includegraphics[draft=false,keepaspectratio=true,clip, %
                   width=0.9\linewidth]%
                   {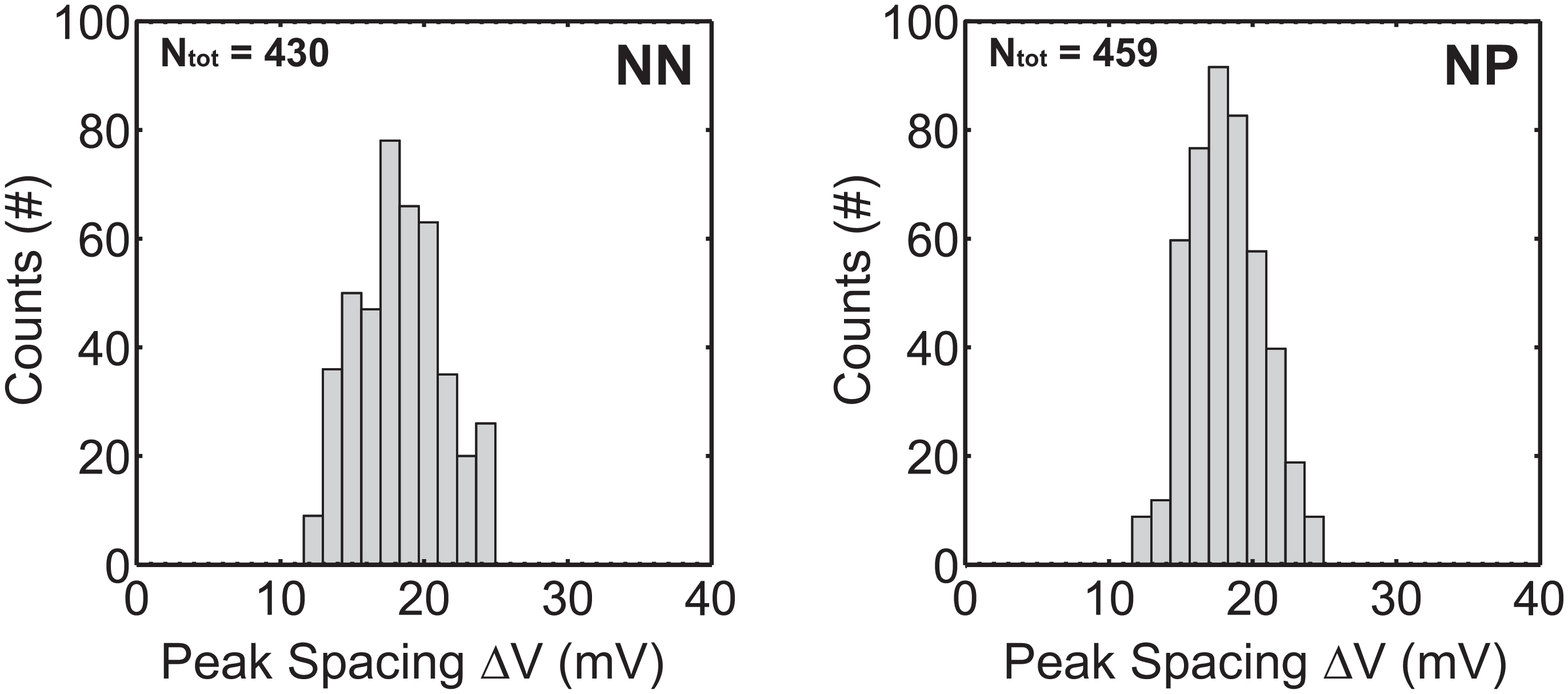}                   
\caption[FIG5]{
Nearest neighbor Coulomb peak spacing statistics in both the NN (left) and NP (right)
regime. Several plunger gate sweeps (at different barrier gate potentials) have been
evaluated and in total 431 Coulomb peaks in the NN (and 460 peaks in the NP) regime have
been considered for the statistics.} 
\label{trdansport}
\end{figure}
All lever arms and capacitances are summarized in Tab.~1.
It shows that the island geometry and dot location with respect to the lateral
gates stays almost constant.
However, the capacitive coupling to the source and drain contacts (i.e. $C_S$ and $C_D$) changes significantly as function of the tunnel barrier configuration.
This can be nicely seen when comparing the symmetry of the diamonds in the NN and NP regime as shown in Figs.~4e and~4b.
While the size and fluctuations of the diamonds remain (almost) constant the lever arms of the source and drain contacts change
strength. In one case (NP regime) we extract $C_S~\approx~1.8$~aF and $C_D~\approx~9.6$~aF, whereas in the other (NN regime) $C_S~\approx~10.1$~aF and $C_D~\approx~1.8$~aF, which can be seen from the different slopes of the diamond edges. However, the individual tunnel barriers strongly depend on the local barrier configuration and change also within the NN or the NP region.

We now estimate the energy scale of the resonances in the
constrictions. The spacing of the constriction resonances in plunger gate is
about 200~mV, whereas the spacing of Coulomb peaks is 20~mV. 
By assuming that the capacitance between the plunger gate and the localized states in the constrictions leading to the resonances is about three times smaller than $C_{pg}$ (estimated from the geometry of the device)
the
energy scale of the resonances in the constriction is about 10~mV, in
agreement with the measured gap in Fig.~2f.

Alternatively, this characteristic energy scale can also be estimated by
considering that the back gate voltage sweep from -25~V to -15~V (around the charge
neutrality point at $V_{bg}~=~$-20~V, Fig.~1c) translates to a Fermi energy
sweep over an energy interval of approx. 120~meV. Near the Dirac point the 
spacing of the constriction resonances in back gate voltage is found to be of
the order of 200~mV leading again to a characteristic energy scale of 10~meV. 

%---------------
Finally, we also performed Coulomb peak spacing ($\Delta V$) statistics in both, the NN and NP regime with in total more than 900 Coulomb peaks, as shown in Fig.~5.
The mean nearest neighbor spacing of the Coulomb peaks in both the NN and NP regime do not differ significantly ($\overline{\Delta V}_{NN}$=17.4~mV and 
$\overline{\Delta V}_{NP}$=17.9~mV). The broadening of the peak spacing distribution is in both cases significant ranging from $\sigma_{NN} \approx$~3.3mV (0.6~meV) to $\sigma_{NP} \approx$~2.5mV (0.5~meV), which is in agreement with Ref.~\cite{pon08}. 
The broadening of the observed unimodal peak spacing distribution is significantly larger than the difference between the average spacings in the NN and NP regimes (0.075~meV).
%, and it is of the same order of magnitude as the estimated single level spacing~\cite{com03}.
The inhomogeneity of the island, as indicated by the different slopes in Fig.~3d may significantly contribute to the observed broadening, which might be also partly %
%However, there are no systematic peak spacing fluctuations
% what does this tell us? Can we make a more explicit statement?
%and part of the observed broadening may be 
influenced by the underlying modulation of the transmission through the narrow graphene constrictions. The broadening of the distributions is significantly larger than expected for a purely metallic SET~\cite{for00}. On the other hand the width of the distribution is of the order of the estimated single-particle level spacing~\cite{com03}, similar to previous observations in high-quality GaAs quantum dots~\cite{Patel98,Luscher01}. This may indicate the importance of quantization effects.
%%%%%%%%%%%%%%%%%%%%%%%%
% What is missing in this discussion: It can be seen in Fig. 3(d) that the peak spacing fluctuations are essentially due to anticrossings of states on which the barrier gates B1 and B2 have different lever arms. This is evidence for states localized in different parts of the dot, similar to a strongly coupled double dot system.
%---------------
%---------------
% Comparison to GaAs
%---------------
%---------------

In conclusion, we have fabricated and characterized a fully tunable graphene single electron transistor based on an etched width-modulated graphene nanostructure with lateral graphene gates. Its functionality was demonstrated by observing electrostatic control over the tunneling barriers. From Coulomb diamond measurements it was estimated that the charging energy of the graphene island is $\approx$~3.4~meV, compatible with its lithographic dimensions. 
These results give detailed insights into tunable graphene quantum dot devices and open the way to  study graphene quantum dots with smaller dimensions and at lower temperatures.

{Acknowledgment ---}
The authors wish to thank R.~Leturcq, P.~Studerus, C.~Barengo, P.~Strasser, A.~Castro-Neto and K.~S.~Novoselov for helpful discussions. Support by the ETH FIRST Lab and financial support by the Swiss National Science Foundation and NCCR nanoscience are gratefully acknowledged.


\begin{thebibliography}{99}

\bibitem{nov04}
K. S. Novoselov, A. K. Geim, S. V. Morozov, D.~Jiang, M.~I.~Katsnelson, S.~V.~Dubonos, I.~V.~Grigorieva, A.~A.~Firsov, Science,~{\bf 306}, 666, (2004).

\bibitem{gei07}
For review see: A. K. Geim and K. S. Novoselov, Nat. Mater.~{\bf 6}, 183
(2007), A.~H.~Castro Neto, F.~Guinea, N.~M.~Peres, A.~K.~Geim, cond-mat 0709.1163v1 (2007).

\bibitem{joa00}
C.~Joachim, J.~K.~Gimzewski and A.~Aviram, Nature,~{\bf 408}, 541, (2000).

\bibitem{min06}
H.~Min, J. E. Hill, N. A. Sinitsyn, B.~R.~Sahu, L.~Kleinman, and A.~H.~MacDonald, Phys. Rev. B,~{\bf 74}, 165310, (2006).
¨
\bibitem{tom07}
N.~Tombros, C.~Jozsa, M.~Popinciuc, H.~T.~Jonkman and B.~J.~van Wees, Nature,~{\bf 448}, 571-574, (2007).

\bibitem{tra07}
B. Trauzettel, D.V. Bulaev, D.~Loss, and G.~Burkard, Nature Physics,~{\bf 3}, 192, (2007).

%\bibitem{kat07}
%M. I. Katnelson, Materials Today,~ {\bf 10(1-2)}, 20-27, (2007).

\bibitem{che07}
Z. Chen, Y. Lin, M. Rooks and P. Avouris, Physica E,~{\bf 40}, 228, (2007).

\bibitem{han07}
M. Y. Han, B. \"Ozyilmaz, Y. Zhang, and P. Kim, Phys. Rev. Lett.,~{\bf 98}, 206805 (2007).

\bibitem{ieo04}
M.~Ieong, B.~Doris, K.~Kedzierski, K.~Rim and M.~Yang, Science,~{\bf 306}, 2057 (2004).

\bibitem{dom99}
N. Dombay, and A. Calogeracos, Phys. Rep.,~{\bf 315}, 41–58 (1999).

\bibitem{kat06}
M. I. Katsnelson, K. S. Novoselov, amd A. K. Geim, Nature Phys.~{\bf 2},
620–625 (2006).

\bibitem{sol07}
F.~Sols, F.~Guinea and A.~H.~Castro Neto, Phys. Rev. Lett.,~{\bf 99}, 166803 (2007).

\bibitem{dai08}

X. Li, X. Wang, L. Zhang, S. Lee, H. Dai, Science,~{\bf 319}, 1229 (2008).

\bibitem{sta08}
C. Stampfer, J. G\"uttinger, F. Molitor, D. Graf, T. Ihn, and K. Ensslin, Appl. Phys. Lett.,~{\bf 92}, 012102 (2008).

\bibitem{pon08}
L. A. Ponomarenko, F. Schedin, M. I. Katsnelson, R.~Yang, E.~H.~Hill, K.~S.~Novoselov, A.~K.~Geim,
Science,~{\bf 320}, 356 (2008).

\bibitem{mia07}
F.~Miao, S.~Wijeratne, Y.~Zhang, U.~C.~Coskun, W.~Bao, C.~N.~Lau, Science,~{\bf 317}, 1530 (2007).

\bibitem{kou97}
L. P. Kouwenhoven et al. 1997, "Electron transport in quantum dots", Mesoscopic
Electron Transport (ed) L.~L.~Sohn, L.~P.~Kouwenhoven and G. Sch\"on (NATO
Series, Kluwer, Dordrecht).


%\bibitem{nov05}
%K. S. Novoselov, A. K. Geim, S. V. Morozov, D.~Jiang, M.~I.~Katsnelson, I.~V.~Grigorieva, S.~V.~Dubonos, A. A. Firsov,  Nature,~{\bf 438}, 197-200, (2005).

%\bibitem{zha05}
%Y. Zhang, Y.-W.~Tan, H.~L.~Stormer, P.~Kim, Nature,~{\bf 438}, 201-204, (2005).

%\bibitem{bun05} 
%J. S. Bunch, Y. Yaish, M. Brink, K. Bolotin, and P. L. McEuen, Nano Lett.,~{\bf 5~(2)}, 287-290, (2005).




%\bibitem{sta07a}
%C. Stampfer, A. B\"urli, A. Jungen, and C. Hierold,
%Phys. Stat. Sol. (B), 1–5 / DOI 10.1002 (2007).

\bibitem{fer06} 
A.~C. Ferrari, J.~C. Meyer, V. Scardaci, C. Casiraghi, M. Lazzeri, F. Mauri,
S. Piscanec, D. Jiang, K.~S. Novoselov, S. Roth, and A.~K. Geim,
Phys. Rev. Lett. {\bf 97}, 187401 (2006).

\bibitem{dav07a} D. Graf, F. Molitor, K. Ensslin, C. Stampfer, A. Jungen, C. Hierold, and
L. Wirtz, Nano Lett., {\bf 7}, 238 (2007).

\bibitem{mol07}
F. Molitor, J. G\"uttinger, C. Stampfer, D. Graf, T. Ihn, and
K. Ensslin,  Phys. Rev. B {\bf 76}, 245426 (2007).

%\bibitem{gupta}
%A. Gupta, G. Chen, P. Joshi, S.~Tadigadapa, and P.~C.~Eklund,
%Nano Lett., {\bf 6}, 2667 (2006).

%\bibitem{com00}
%This can be best seen in Figs.~2a and 2b. At a back gate voltage of approx.~17.5V the saturation
%current is significantly lifted and does not depend on the barrier gate voltages 
%$V_{b1}$ and $V_{b2}$. Therefore this current is limited outside the barriers and island; thus, in the contact regions.

\bibitem{com01}
Here an offset of 5~V has been used in order to compensate the slightly different doping
of the two narrow graphene constrictions. This offset can also be nicely seen at $V_{b2}$ in Fig.~2c.

\bibitem{ihn04}
See Fig.~12.11 in T. Ihn, Springer Tracts in Modern Physics 192, Springer 2004, p. 102.


\bibitem{for00}
M. Furlan, T. Heinzel, B. Jeanneret, S. V. Lotkhov and K. Ensslin,
Europhys. Lett.,~{\bf 49}, 369 (2000).

\bibitem{com03}
The single level spacing is estimated by $\Delta_{qm}=\hbar v_F / 2 \sqrt{A N/\pi}$, where $A$ is the
area of the graphene island and $N$ is the number of carriers on the island. For $N$=10 to 100 electrons
on the island we estimate a characteristic energy spacing of 0.75~meV to 0.24~meV.

\bibitem{Patel98} 
S. R. Patel, S. M. Cronenwett, D. R. Stewart, A. G. Huibers, C. M. Marcus, C. I. Duru\"oz and J. S. Harris, Jr., K. Campman and A. C. Gossard, Phys. Rev. Lett.~{\bf 80}, 4522 - 4525 (1998)

\bibitem{Luscher01} S. L\"uscher, T. Heinzel, K. Ensslin, W. Wegscheider, and M. Bichler, Phys. Rev. Lett.~{\bf 86}, 2118 (2001)

%\bibitem{bee90}
%C. W. J. Beenakker, Phys. Rev. B,~{\bf 44}, 1646 (1990).

%\bibitem{grb05}
%B. Grbic, R. Leturcq, K. Ensslin, D. Reuter, and A. D.~Wiek, Appl. Phys. Lett.,~{\bf 87}, 232108, (2005)



\end{thebibliography}
\end{document}